\documentclass[conference]{IEEEtran}
\IEEEoverridecommandlockouts
\usepackage{cite}
\usepackage{amsmath,amssymb,amsfonts}
\usepackage{algorithmic}
\usepackage{graphicx}
\usepackage{textcomp}
\usepackage{xcolor}
\usepackage{makecell}
\usepackage{verbatim}
\usepackage{cuted}
\usepackage{amsmath}\allowdisplaybreaks
\def\BibTeX{{\rm B\kern-.05em{\sc i\kern-.025em b}\kern-.08em
    T\kern-.1667em\lower.7ex\hbox{E}\kern-.125emX}}
\usepackage[letterpaper, top=0.75in, bottom=1.1in, left=0.65in, right=0.65in]{geometry}

\begin{document}

\title{Distributed Indirect Source Coding with Decoder Side Information \\

}

\author{\IEEEauthorblockN{Jiancheng Tang\IEEEauthorrefmark{1}, Qianqian Yang\IEEEauthorrefmark{1}, Deniz Gündüz\IEEEauthorrefmark{2}}
\IEEEauthorblockA{\IEEEauthorrefmark{1}College of information Science and Electronic Engineering, Zhejiang University, Hangzhou 310007, China\\
\IEEEauthorrefmark{1}Email: \{jianchengtang, qianqianyang20\}@zju.edu.cn
\\\IEEEauthorrefmark{2}Department of Electrical and Electronic Engineering Imperial College London London, UK\\
\IEEEauthorrefmark{2}Email:d.gunduz@imperial.ac.uk}
\thanks{{\thefootnote}{*}This work is partly supported by NSFC under grant No. 62293481, No. 62201505, partly by the SUTD-ZJU IDEA Grant (SUTD-ZJU (VP) 202102).}
}

\maketitle
\begin{abstract}
This paper studies a variant of the rate-distortion problem  motivated by task-oriented semantic communication and distributed learning problems, where $M$ correlated sources are independently encoded for a central decoder. The decoder has access to a correlated side information in addition to the messages received from the encoders, and aims to recover a latent random variable correlated with the sources observed by the encoders within a given distortion constraint rather than recovering the sources themselves. We provide bounds on the rate-distortion region for this scenario in general, and characterize the rate-distortion function exactly when the sources are conditionally independent given the side information.

\begin{IEEEkeywords} Semantic communication, distributed source coding, rate-distortion theory, side information.
\end{IEEEkeywords}
\end{abstract}

\section{Introduction}

\begin{figure}
\centerline{\includegraphics[width=3in]{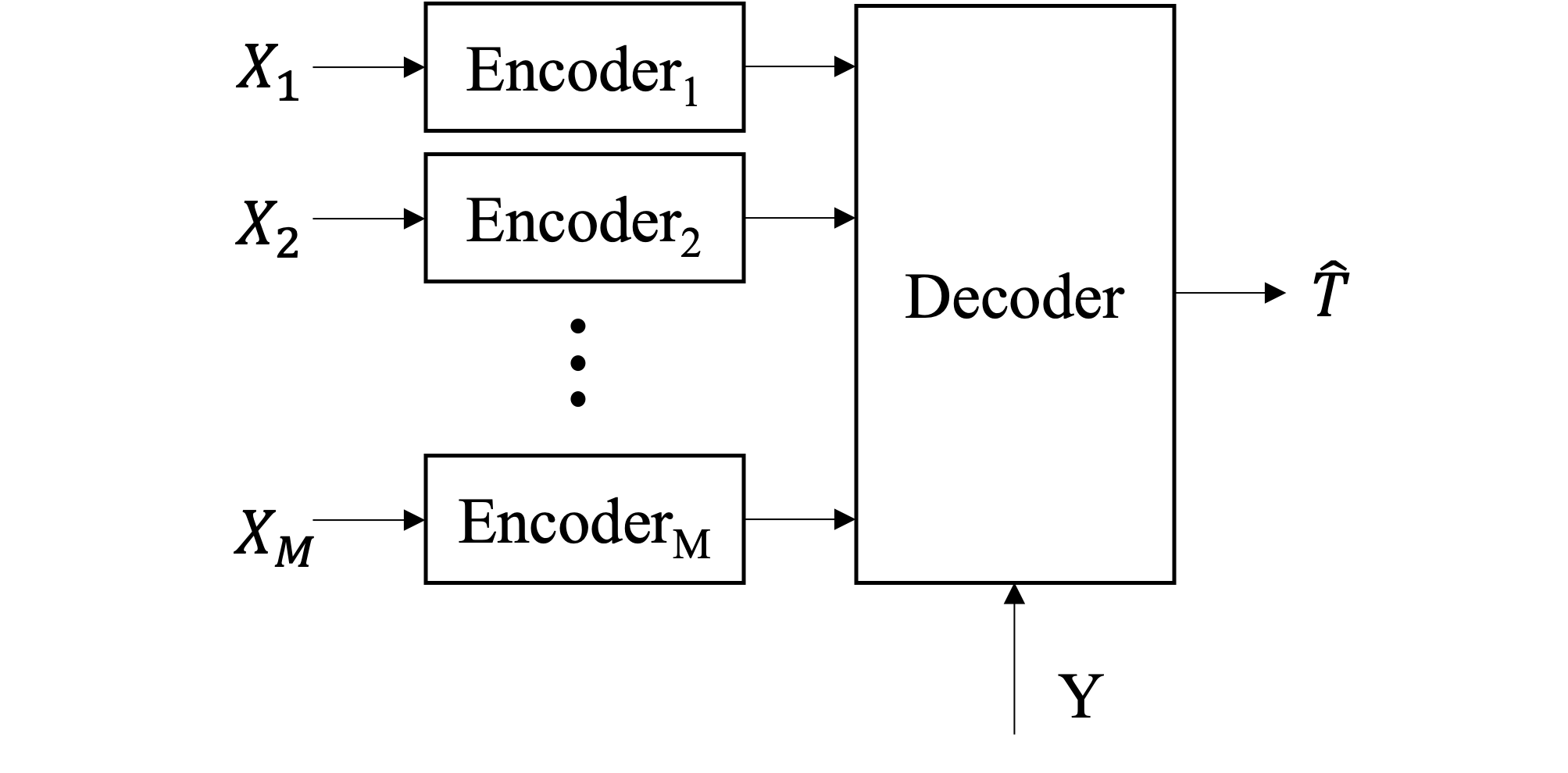}}
\caption{Distributed remote compression of a latent variable by $M$ correlated sources with side information at the receiver.}
\label{figsystem}
\end{figure}

Consider the multiterminal source coding setup as shown in Fig.~\ref{figsystem}. Let $(T, X_1,...,X_M,Y)\sim p(t,x_1,...,x_M,y)$ be a discrete memoryless source (DMS) taking values in the finite alphabets $\mathcal{T} \times \mathcal{X}_1 \times \cdots  \times \mathcal{X}_M \times \mathcal{Y}$ according to a fixed and known probability distribution $p(t,x_1,...,x_M,y)$. In this setup, the encoder $m, m\in \mathcal{M}:=\{1,...,M\}$ has local observations 
${X}_m^n:=(X^n_1, \ldots, X^n_M)$. The agents independently encode their observations into binary sequences at rates $\{R_1,\ldots,R_M\}$ bits per input symbol, respectively. The decoder with side information $Y^n = (Y_1, \ldots, Y_n)$ aims to recover some task-oriented latent information ${T}^n:=(T_1, \ldots, T_n)$ which is correlated with $(X^n_1, \ldots, X^n_M)$, but it is not observed directly by any of the encoders. We are interested in the lossy reconstruction of $T^n$ with the average distortion measured by $\mathbb{E}\left[\frac{1}{n}\sum_{i=1}^n d(T_i,\hat{T}_i) \right]$, for some prescribed single-letter distortion measure $d(\cdot, \cdot)$.  A formal $(2^{n R_1},...,2^{n R_M}, n)$ rate-distortion code for this setup consists of
\begin{itemize}
    \item $M$ independent encoders, where encoder $m \in \mathcal{M}$ assigns an index $s_m(x_m^n) \in \left\{1, \ldots, 2^{n R_m}\right\}$ to each sequence $x_m^n \in \mathcal{X}_m^n$;
    \item a decoder that produces the estimate  $\hat{t}^n(s_1,...,s_M,y^n) \in \mathcal{T}^n$ to each index tuple $(s_1,...,s_M)$ and side information $y^n \in \mathcal{Y}^n$.
\end{itemize}

A rate tuple $(R_1,...,R_M)$ is said to be achievable with the distortion measure $d(\cdot, \cdot)$ and the distortion value $D$ if there exists a sequence of $(2^{n R_1},...,2^{n R_M}, n)$ codes that satisfy
\begin{equation}
\begin{aligned}
\mathop {\lim \sup }\limits_{n \to \infty } \mathbb{E}\left[\frac{1}{n}\sum_{i=1}^n d(T_i,\hat{T}_i) \right] \le D.
\end{aligned}
\label{distortion constraint}
\end{equation}

The rate-distortion region $R_{{X_1},\ldots,{X_m}|Y}^{*}\left( {{D}{}} \right)$ for this distributed source coding problem is the closure of the set of all achievable rate tuples $(R_1,\ldots,R_M)$ that permit the reconstruction of the latent variable $T^n$ within the average distortion constraint $D$.


The problem as illustrated in Fig.~\ref{figsystem} is motivated by semantic/ task-oriented communication and distributed learning problems. In semantic/task-oriented communication, the decoder only needs to reconstruct some task-oriented information implied by the sources. For instance, it might extract hidden features from a scene captured by multiple cameras positioned at various angles. Here, $T_i$ may also be a deterministic function of the source samples $(X_{1,i}, \ldots, X_{M,i})$, which then reduces to the problem of lossy distributed function computation \cite{two, howto, distributed}. A similar problem also arises in distributed training. Consider $Y^n$ as the global model available at the server at an iteration of a federated learning process, and $(X^n_1, \ldots, X^n_M)$ as the independent correlated versions of this model after downlink transmission and local training. The server aims to recover the updated global model, $T^n$, based on the messages received from all $M$ clients. It is often assumed that the global model is transmitted to the clients intact, but in practical scenarios where downlink communication is limited, the clients may receive noisy or compressed versions of the global model \cite{amiri2020federated, gruntkowska2024improving, Amiri:TWC:22}. 

For the case of $M=1$, the considerd problem reduces to the remote compression in a point-to-point scenario with side information available at the decoder. In \cite{Photios1, liu}, the authors studied this problem without the correlated side information at the receiver, motivated in the context of semantic communication. This problem is known in the literature as the remote rate-distortion problem \cite{Dobrushin, Wolf:TIT:70}, and the rate-distortion trade-off is fully characterized in the general case. The authors studied this trade-off in detail for specific source distributions in \cite{Photios1}. Similarly, the authors of \cite{Guo} characterized the remote rate-distortion trade-off when correlated side information is available both at the encoder and decoder. Our problem for $M=1$ can be solved by combining the remote rate-distortion problem with the classical Wyner-Ziv rate-distortion function \cite{wyner1, wyner2}.

The rate-distortion region for the multi-terminal version of the remote rate-distortion problem considered here remains open. Sung \emph{et al.} proposed an achievable rate region for the distributed lossy computation problem, but no conclusive rate-distortion function can be given \cite{Sung}. Gwanmo \emph{et al.} considered a special case in which the sources are independent and derived a single-letter expression for the rate-distortion region \cite{Gwanmo}. Gastpar \cite{Michael} considered the lossy compression of the source sequences in the presence of side information at the receiver. He characterized the rate-distortion region for the special case, in which $X_i$'s are conditionally independent given the side information.


In this paper, we are interested in the rate-distortion region $R_{{X_1},\ldots,{X_m}|Y}^{*}\left( {{D}{}} \right)$ for the general problem. We will pay particular attention to the special case in which the sources are conditionally independent given the side information, motivated by the aforementioned examples. For the sake of brevity of the presentation, we set $M=3$ in this paper, with the understanding that the results can be readily extended to an arbitrary number of sources.

In Section II, we derive an achievable region $R_a\left( {{D}} \right) \subseteq R_{{X_1},{X_2},{X_3}|Y}^{*}\left( {{D}} \right)$.
In Section III, we determine a general outer bound $R_o\left( {{D}} \right)  \supseteq  R_{{X_1},{X_2},{X_3}|Y}^{*}\left( {{D}} \right)$.
In Section IV, we show that the two regions coincide and the region is optimal when the sources $(X_1, X_2,X_3)$ are conditionally independent given the side information $Y$.

\section{An Achievable Rate Region}

In this section, we introduce an achievable rate region $R_a\left( {D} \right) $, which is contained within the goal rate-distortion region $R_a\left( {{D}} \right) \subseteq R_{{X_1},{X_2},{X_3}|Y}^{*}\left( {{D}} \right)$.

\emph{Theorem 1:} $R_a\left( {{D}} \right) \subseteq R_{{X_1},{X_2},{X_3}|Y}^{*}\left( {{D}} \right)$, where $R_a\left( {{D}} \right)$ is the set of all rate tuples $(R_1,R_2,R_3)$ such that there exists a tuples  $(W_1,W_2,W_3)$ of discrete random variables with $p\left( {{w_1},{w_2},{w_3},{x_1},{x_2},{x_3},y} \right) = p\left( {{w_1}|{x_1}} \right)p\left( {{w_2}|{x_2}} \right)p\left( {{w_3}|{x_3}} \right)p\left( {{x_1},{x_2},{x_3},y} \right)$, for which the following conditions are satisfied
\begin{subequations} \label{achieve region11}
\begin{small}
\begin{align}
{R_1} &\geqslant I\left( {{X_1};{W_1}} \right) - I\left( {{W_1};{W_2},{W_3},Y} \right) \hfill \label{2a}\\
  {R_2} &\geqslant I\left( {{X_2};{W_2}} \right) - I\left( {{W_2};{W_1},{W_3},Y} \right) \hfill \label{2b}\\
  {R_3} &\geqslant I\left( {{X_3};{W_3}} \right) - I\left( {{W_3};{W_1},{W_2},Y} \right) \hfill \label{2c} \\
  {R_1} + {R_2} &\geqslant I\left( {{X_1};{W_1}} \right) + I\left( {{X_2};{W_2}} \right) - I\left( {{W_1};{W_2},{W_3},Y} \right) \nonumber \\&- I\left( {{W_2};{W_1},{W_3},Y} \right) + I\left( {{W_1};{W_2}|{W_3},Y} \right) \hfill \label{2d}\\
  {R_1} + {R_3} &\geqslant I\left( {{X_1};{W_1}} \right) + I\left( {{X_3};{W_3}} \right) - I\left( {{W_1};{W_2},{W_3},Y} \right) \nonumber\\&- I\left( {{W_3};{W_1},{W_2},Y} \right) + I\left( {{W_1};{W_3}|{W_2},Y} \right) \hfill \label{2e} \\
 {R_2} + {R_3} &\geqslant I\left( {{X_2};{W_2}} \right) + I\left( {{X_3};{W_3}} \right) - I\left( {{W_2};{W_1},{W_3},Y} \right) \nonumber \\&- I\left( {{W_3};{W_1},{W_2},Y} \right) + I\left( {{W_2};{W_3}|{W_1},Y} \right) \hfill \label{2f},\\   {R_1} + {R_2} + {R_3} &\geqslant I\left( {{X_1};{W_1}} \right) + I\left( {{X_2};{W_2}} \right) + I\left( {{X_3};{W_3}} \right) \nonumber\\&- I\left( {{W_1};{W_2},{W_3},Y} \right) - I\left( {{W_2};{W_1},{W_3},Y} \right) \nonumber\\&- I\left( {{W_3};{W_1},{W_2},Y} \right) + I\left( {{W_1};{W_2}|{W_3},Y} \right)\nonumber \\&+ I\left( {{W_1},{W_2};{W_3}|Y} \right) \label{2g},
\end{align}
\end{small}
\end{subequations}

and there exist a decoder ${g}\left(  \cdot  \right)$  such that
\begin{equation}
\begin{aligned}
E{d}({T},{g(W_1,W_2,W_3,Y)}) \leqslant {D}.\\
\end{aligned}
\label{constrains}
\end{equation}

The rigorous proof of Theorem 1 is provided in Appendix A.

\emph{Corollary 2:} The conditions \eqref{constrains} of Theorem 1 can be expressed equivalently as
\begin{subequations} \label{corollary2}
\begin{align}
{R_1} &\geqslant I\left( {{X_1},{X_2},{X_3};{W_1}|{W_2},{W_3},Y} \right) \label{corollary2a} \\
{R_2} &\geqslant I\left( {{X_1},{X_2},{X_3};{W_2}|{W_1},{W_3},Y} \right) \label{corollary2b} \\
{R_3} &\geqslant I\left( {{X_1},{X_2},{X_3};{W_3}|{W_1},{W_2},Y} \right) \label{corollary2c}\\
{R_1} + {R_2} &\geqslant I\left( {{X_1},{X_2},{X_3};{W_1},{W_2}|{W_3},Y} \right) \label{corollary2d} \\
{R_1} + {R_3} &\geqslant I\left( {{X_1},{X_2},{X_3};{W_1},{W_3}|{W_2},Y} \right) \\
{R_2} + {R_3} &\geqslant I\left( {{X_1},{X_2},{X_3};{W_2},{W_3}|{W_1},Y} \right) \\
{R_1} + {R_2} + {R_3} &\geqslant I(X_1, X_2, X_3; W_1, W_2, W_3| Y)
\end{align}
\end{subequations}

\emph{Proof:} First we prove ${R_1} \geqslant I\left( {{X_1};{W_1}} \right) - I\left( {{W_1};{W_2},{W_3},Y} \right) = I\left( {{X_1},{X_2},{X_3};{W_1}|{W_2},{W_3},Y} \right)$. The bound of \eqref{corollary2a} can be written as
\begin{equation}
\begin{aligned}
    &I\left( {{X_1},{X_2},{X_3};{W_1}|{W_2},{W_3},Y} \right) \\
    &= I\left( {{X_1};{W_1}|{W_2},{W_3},Y} \right)   + \underbrace {I\left( {{X_2},{X_3};{W_1}|{X_1},{W_2},{W_3},Y} \right)}_{=0},
\end{aligned}
\label{5}
\end{equation}
where $I\left( {{X_2},{X_3};{W_1}|{X_1},{W_2},{W_3},Y} \right) = 0 $ because $(X_2, X_3, Y)$ is conditionally independent of $W_1$ for given $X_1$. For the first term of the right in \eqref{5}, we have 
\begin{equation}
\begin{aligned}
  &I\left( {{X_1};{W_1}|{W_2},{W_3},Y} \right) + I\left( {{W_1};{W_2},{W_3},Y} \right) \\
  &= I\left( {{W_1};{X_1},{W_2},{W_3},Y} \right) \hfill \\
   &= I\left( {{W_1};{X_1}} \right) + \underbrace {I\left( {{W_2},{W_3},Y;{W_1}|{X_1}} \right)}_{=0},
\end{aligned}
\end{equation}
where $I\left( {{W_2},{W_3},Y;{W_1}|{X_1}} \right)=0$ because $(W_2,W_3,Y)$ is conditionally independent of $W_1$ given $X_1$. Then we have
\begin{equation}
\begin{aligned}
&I\left( {{X_1},{X_2},{X_3};{W_1}|{W_2},{W_3},Y} \right) \\
&= I\left( {{X_1};{W_1}|{W_2},{W_3},Y} \right) \\
&= I\left( {{W_1};{X_1}} \right) - I\left( {{W_1};{W_2},{W_3},Y} \right)\\
& \le R_1 .
\end{aligned}
\end{equation}
This completes the proof of \eqref{corollary2a}. \eqref{corollary2b} and \eqref{corollary2c} can be proved in the same way. The rigorous proof of the rest sum rate bounds will be provided in a longer version.

\section{An General Outer Bound}
In this section, we derive a region $R_o\left( {{D}} \right) $ which contains the goal rate-distortion region $R_o\left( {{D}} \right)  \supseteq  R_{{X_1},{X_2},{X_3}|Y}^{*}\left( {{D}} \right)$.

\emph{Theorem 3:} $R_o\left( {{D}} \right)  \supseteq  R_{{X_1},{X_2},{X_3}|Y}^{*}\left( {{D}} \right)$, where $R_o\left( {{D}} \right) $ is the set of all rate triples $(R_1,R_2,R_3)$ such that there exists a triple  $(W_1,W_2,W_3)$ of discrete random variables with  $p\left( {{w_1}|{x_1},{x_2},{x_3}, y} \right) = p\left( {{w_1}|{x_1}} \right)$, $p\left( {{w_2}|{x_1},{x_2},{x_3}, y} \right) = p\left( {{w_2}|{x_2}} \right)$ and $p\left( {{w_3}|{x_1},{x_2},{x_3}, y} \right) = p\left( {{w_3}|{x_3}} \right)$,  for which the following conditions are satisfied
\begin{equation}
\begin{small}
\begin{aligned}
{R_1} &\geqslant I\left( {{X_1},{X_2},{X_3};{W_1}|{W_2},{W_3},Y} \right) \\
{R_2} &\geqslant I\left( {{X_1},{X_2},{X_3};{W_2}|{W_1},{W_3},Y} \right) \\
{R_3} &\geqslant I\left( {{X_1},{X_2},{X_3};{W_3}|{W_1},{W_2},Y} \right) \\
{R_1} + {R_2} &\geqslant I\left( {{X_1},{X_2},{X_3};{W_1},{W_2}|{W_3},Y} \right) \\
{R_1} + {R_3} &\geqslant I\left( {{X_1},{X_2},{X_3};{W_1},{W_3}|{W_2},Y} \right) \\
{R_2} + {R_3} &\geqslant I\left( {{X_1},{X_2},{X_3};{W_2},{W_3}|{W_1},Y} \right) \\
{R_1} + {R_2} + {R_3} &\geqslant I(X_1, X_2, X_3; W_1, W_2, W_3| Y)
\end{aligned}
\label{outerbound}
\end{small}
\end{equation}

and there exist a decoding function ${g}\left(  \cdot  \right)$ such that

\begin{equation}
\begin{aligned}
E{d}({T},{g_1(W_1,W_2,W_3,Y)}) \leqslant {D},\\
\end{aligned}
\label{constrains}
\end{equation}

The rigorous proof of Theorem 3 is provided in Appendix B.

While the expressions of the inner bound \eqref{corollary2} and the outer bound \eqref{outerbound} are the same, these two regions do not coincide because the marginal constrains $p\left( {{w_1},{w_2},{w_3},{x_1},{x_2},{x_3},y} \right) = p\left( {{w_1}|{x_1}} \right)p\left( {{w_2}|{x_2}} \right)p\left( {{w_3}|{x_3}} \right)p\left( {{x_1},{x_2},{x_3},y} \right)$ in Theorem 1 limit the degree of freedom for choosing the auxiliary random variables $(W_1, W_2, W_3)$ compared with the marginal constrains in Theorem 3. In the next section, we will demonstrate that the additional degree of freedom in choosing the auxiliary random variables $(W_1, W_2, W_3)$ in Theorem 3 cannot lower the value of the rate-distortion functions.


\section{conclusive rate-distortion results}

\emph{Corollary 4:} If $X_1, X_2, X_3$ are conditionally independent given the side information $Y$, $R_a\left( {{D}} \right) \subseteq R_{{X_1},{X_2},{}|Y}^{*}\left( {{D}} \right)$ where $R_a\left( {{D}} \right) $ is the set of all rate triples $(R_1,R_2,R_3)$ such that there exists a triple $(W_1,W_2,W_3)$ of random variables with $p\left( {{w_1},{w_2},{w_3},{x_1},{x_2},{x_3},y} \right) = p\left( {{w_1}|{x_1}} \right)p\left( {{w_2}|{x_2}} \right)p\left( {{w_3}|{x_3}} \right)p\left( {{x_1}|y} \right) p\left( {{x_2}|y} \right) p\left( {{x_3}|y} \right)  p\left( {y} \right)$, for which the following conditions are satisfied
\begin{equation}
\begin{aligned}
  {R_1} &\geqslant I\left( {{X_1};{W_1}} \right) - I\left( {{W_1};Y} \right) \hfill \\
  {R_2} &\geqslant I\left( {{X_2};{W_2}} \right) - I\left( {{W_2};Y} \right) \hfill \\
    {R_3} &\geqslant I\left( {{X_3};{W_3}} \right) - I\left( {{W_3};Y} \right) \hfill \\
\end{aligned}
\label{achieve region16}
\end{equation}
and there exist decoding functions ${g}\left(  \cdot  \right)$ such that

\begin{equation}
\begin{aligned}
E{d}({T},{g(W_1,W_2,W_3,Y)}) \leqslant {D}.\\
\end{aligned}
\label{constrains}
\end{equation}

\emph{Proof:} Since the joint distribution can be written as 
\begin{equation}
\begin{aligned}
&p\left( {{w_1},{w_2},{w_3},{x_1},{x_2},{x_3},y} \right)
\\& =p\left( {{w_1},{w_2},{w_3}|{x_1},{x_2},{x_3},y} \right)p\left( {{x_1},{x_2},{x_3}|y} \right)  p\left( {y} \right) \\& =
p\left( {{w_1}|{x_1}} \right)p\left( {{w_2}|{x_2}} \right)p\left( {{w_3}|{x_3}} \right)p\left( {{x_1}|y} \right)\\& \times p\left( {{x_2}|y} \right) p\left( {{x_3}|y} \right)  p\left( {y} \right),
\end{aligned}
\label{distribution}
\end{equation}
the terms $I\left( {{W_1};{W_2}|{W_3},Y} \right)$ in the sum rate bound \eqref{2d} is $0$  because $W_2$ is conditionally independent of $W_1$. Similarly, the terms $I\left( {{W_1};{W_3}|{W_2},Y}\right), I\left( {{W_2};{W_3}|{W_1},Y} \right)$ and $I\left( {{W_1};{W_2}|{W_3},Y} \right)+$$I\left( {{W_1},{W_2};{W_3}|Y} \right)$ in the sum rate bound \eqref{2e}-\eqref{2g} are all $0$. Therefore, the sum rate bound can be expressed as the combination of the side bounds, and hence can be omitted. 
Meanwhile,  The term $I\left( {{W_1};{W_2},{W_3},Y} \right) $ in the side bound in \eqref{2a} can be written as
\begin{equation}
\begin{aligned}
I\left( {{W_1};{W_2},{W_3},Y} \right) = I\left( {{W_1};Y} \right) + \underbrace {I\left( {{W_2},{W_3};{W_1}|Y} \right)}_{=0}. \\
\end{aligned}  
\end{equation}
Similarly, we have
\begin{equation}
\begin{aligned}
I\left( {{W_2};{W_1},{W_3},Y} \right) = I\left( {{W_2};Y} \right) + \underbrace {I\left( {{W_1},{W_3};{W_2}|Y} \right)}_{=0} \\
I\left( {{W_3};{W_1},{W_2},Y} \right) = I\left( {{W_3};Y} \right) + \underbrace {I\left( {{W_1},{W_2};{W_3}|Y} \right)}_{=0}.
\end{aligned}  
\end{equation}
This completes the  proof  Corollary 4.


\emph{Corollary 5:} If $X_1, X_2, X_3$ are conditionally independent given the side information $Y$, $R_o^{'}\left( {{D}} \right)  \supseteq  R_o\left( {{D}} \right)$, and hence  $R_o^{'}\left( {{D}} \right)  \supseteq  R_{{X_1},{X_2},{W_3}|Y}^{*}\left( {{D}} \right)$ where $R_o^{'}\left( {{D}} \right)$ is the set of all rate triples $(R_1,R_2,R_3)$ such that there exists a triple $(W_1,W_2,W_3)$ of discrete random variables with $p\left( {{w_1}|{x_1},{x_2},{w_3},y} \right) = p\left( {{w_1}|{x_1}} \right)$, $p\left( {{w_2}|{x_1},{x_2},{w_3},y} \right) = p\left( {{w_2}|{x_2}} \right)$ and $p\left( {{w_3}|{x_1},{x_2},{x_3},y} \right) = p\left( {{w_3}|{x_3}} \right)$, for which the following conditions are satisfied

\begin{equation}
\begin{small}
\begin{aligned}
\small
  {R_1} &\geqslant I\left( {{X_1};{W_1}} \right) - I\left( {{W_1};Y} \right) \hfill \\
  {R_2} &\geqslant I\left( {{X_2};{W_2}} \right) - I\left( {{W_2};Y} \right)
\\{R_3} &\geqslant I\left( {{X_3};{W_3}} \right) - I\left( {{W_3};Y} \right) \hfill ,
\normalsize
\end{aligned}
\label{outer bound}
\end{small}
\end{equation}
and  there exists decoding functions ${g}\left(  \cdot  \right)$ such that
\begin{equation}
\begin{aligned}
E{d}({T},{g_1(W_1,W_2,W_3,Y)}) \leqslant {D},
\end{aligned}
\label{constrains}
\end{equation}
\emph{Proof:} First, we can enlarge the region $R_o\left( {{D}} \right)$ by omitting the sum rate bound in \eqref{outerbound}. Then, the side rate bounds in \eqref{outerbound} can be relaxed as
\begin{equation}
\begin{small}
\begin{aligned}
{R_1} &\geqslant I\left( {{X_1},{X_2},{X_3};{W_1}|{W_2},{W_3},Y} \right) \\& \geqslant I\left( {{X_1};{W_1}|{W_2},{W_3},Y} \right) +I\left( {{X_2},{X_3};{W_1}|{W_2},{W_3},Y,X_1} \right) 
\\& \geqslant I\left( {{X_1};{W_1}|{W_2},{W_3},Y} \right)
\\ &= I\left( {{X_1};{W_1},{W_2},{W_3}|Y} \right)-\underbrace {I\left( {{X_1};{W_2},{W_3}|Y} \right)}_{=0}.
\end{aligned} 
\end{small}
\label{proof3}
\end{equation}

According to the conditional independence relations, we have $I\left( {{X_1};{W_2},{W_3}|Y} \right)=0$, and then we have 
\begin{subequations} \label{proof5}
\begin{small}
\begin{align}
{R_1} &\geqslant I\left( {{X_1};{W_1},{W_2},{W_3}|Y} \right) \nonumber \\&= I\left( {{X_1};{W_1}|Y} \right) + \underbrace {I\left( {{X_1};{W_2},{W_3}|Y,{W_1}} \right)}_{=0} \label{24a}
\\& = I\left( {{X_1},Y ;{W_1}} \right) - I\left( {{W_1};Y} \right) \label{24b}
\\& = I\left( {{X_1};{W_1}} \right) - I\left( {{W_1};Y} \right) \label{24c}
\end{align} 
\end{small}
\end{subequations}
where \eqref{24a} is obtained by the condition that $X_1, X_2, X_3$ are conditionally independent given the side information $Y$, \eqref{24b} follows from the  chain rule of mutual information and \eqref{24c} is derived by the Markov chain $Y - X_1 - W_1$.
The same derivation can be applied to $R_2$ and $R_3$, this proves the corollary 5.

\emph{Theorem 6:} If  $X_1, X_2, X_3$ are conditionally independent given the side information $Y$,
\begin{equation}
\begin{aligned}
R_a\left( {{D}} \right) = R_o\left( {{D}} \right)  =  R_{{X_1},{X_2},{X_3}|Y}^{*}\left( {{D}} \right).
\end{aligned}
\label{constrains}
\end{equation}

\emph{Proof:} We note that the only difference between $R_a\left( {{D}} \right)$ and $R_o\left( {{D}} \right)$ is the degrees of freedom when choosing the auxiliary random variables $(W_1, W_2, W_3)$, and all of the mutual information functions in \eqref{achieve region16} and \eqref{outer bound} only depend on the marginal distribution $(X_1, W_1, Y)$, $(X_2, W_2, Y)$ and $(X_3, W_3, Y)$. Randomly choose a certain rate triple $(R_1, R_2, R_3)$ with a auxiliary random variable triple $(W_1, W_2, W_3)$ meeting the conditions of Corollary 5, the corresponding joint distribution is given in \eqref{distribution}
Then we construct the auxiliary random variables $(W_1^{'}, W_2^{'}, W_3^{'})$ such that
\begin{equation}
\begin{footnotesize}
\begin{aligned}
{p_{W_1^{'}|{X_1}}}\left( {{w_1}|{s_1}} \right) = \sum\limits_{{w_2},{s_2},{w_3},{s_3}} {p\left( {{w_1},{w_2},{w_3}|{s_1},{s_2},{s_3}} \right)} p\left( {{s_2},{s_3}|{s_1}} \right)\\
{p_{W_2^{'}|{X_2}}}\left( {{w_2}|{s_2}} \right) = \sum\limits_{{w_1},{s_1},{w_3},{s_3}} {p\left( {{w_1},{w_2},{w_3}|{s_1},{s_2},{s_3}} \right)} p\left( {{s_1},{s_3}|{s_2}} \right) \\
{p_{W_3^{'}|{X_3}}}\left( {{w_3}|{s_3}} \right) = \sum\limits_{{w_1},{s_1},{w_2},{s_2}} {p\left( {{w_1},{w_2},{w_3}|{s_1},{s_2},{s_3}} \right)} p\left( {{s_1},{s_2}|{s_3}} \right).
\end{aligned}
\end{footnotesize}
\end{equation}
The joint distribution
\begin{equation}
\begin{footnotesize}
\begin{aligned}
&p\left( {{w_1^{'}},{w_2^{'}},{w_3^{'}},{x_1},{x_2},{x_3},Y} \right) \\&= p\left( {w_1^{'}}|{x_1}\right)p\left( {w_2^{'}}|{x_2}\right)p\left( {w_3^{'}}|{x_3}\right)p\left( {{x_1}|{y}} \right)p\left( {{x_2}|{y}} \right)p\left( {{x_3}|y} \right) p\left( {y} \right)
\end{aligned}
\end{footnotesize}
\end{equation}
has the same marginal distributions on $(X_1, W_1, Y)$, $(X_2, W_2, Y)$ and $(X_3, W_3, Y)$. Therefore, the additional degree of freedom for choosing the auxiliary random variables $(W_1, W_2, W_3)$ in Corollary 5 can not lower the value of rate-distortion functions. This proves the Theorem 6. The arguments leading to Theorem 6 indicates that the result extends to the $M$ sources scenario.

\section{Conclusion}
A variant rate-distortion problem  was studied in this paper. We first derived an achievable rate-distortion region of this problem, and subsequently, we derived a general outer bound of the rate-distortion region. We show that the two regions coincide and characterize the rate-distortion function exactly under the scenario that the sources are conditionally
independent given the side information.

\section*{Appendix A}
Here we provide the rigorous proof of Theorem 1. 

\emph{Lemma 7 (Extended Markov Lemma)}: Let
\begin{equation}
\begin{aligned}
&p\left( {{w_1},{w_2},{w_3},{x_1},{x_2},{x_3},y} \right) \\&= p\left( {{w_1}|{x_1}} \right)p\left( {{w_2}|{x_2}} \right)p\left( {{w_3}|{x_3}} \right) p\left( {{x_1},{x_2},{x_3},y} \right).
\end{aligned}
\label{extend mk}
\end{equation}
For a fixed $\left( {{x_1^n},{x_2^n},{x_3^n} y^n} \right) \in A_\epsilon^{*\left( n \right)}$, $w_1^n$ , $w_2^n$ and $w_3^n$ are drown from $p(w_1|x_1)$ , $p(w_2|x_2)$ and $p(w_3|x_3)$, respectively. Then 
\begin{equation}
\begin{aligned}
\mathop {\lim }\limits_{n \to \infty } Pr \{ \left( {w_1^n,w_2^n,w_3^n,{x_1^n},{x_2^n},{x_3^n},y^n} \right) \in A_\epsilon^{*\left( n \right)} \} =1 .
\end{aligned}
\label{extend mk_1}
\end{equation}

\emph{Proof:} We can use the Markov Lemma of joint typicality multiple times to prove this lemma, the details are omitted here.


\emph{Proof of Theorem 1.} For $m = 1,2,3$, fix $p\left( {{w_m}|{x_m}} \right)$ and $g(W_1,W_2,W_3,Y)$ such that the distortion constraint $E{d}({T},{\hat T}) \leqslant {D}$ is satisfied. Calculate $p\left( {{w_m}} \right) = \sum\nolimits_{{x_m}}{p\left( {{x_m}} \right)}p\left( {{w_m}|{x_m}} \right)$.

\emph{Generation of codebooks:} Generate ${2^{nR_m^{{{'}}}}}$ i.i.d codewords ${{\mathbf{w}}_{\mathbf{m}}}^n\left( {{s_m}} \right) \sim \prod\nolimits_{i = 1}^n {p\left( {{w_{m,i}}} \right)} $, and index them by ${s_m} \in \left\{ {1,2,...,{2^{nR_m^{{'}}}}} \right\}$. Provide ${2^{nR_m}}$  random bins with indices ${t_m} \in \left\{ {1,2,...,{2^{nR_m}}} \right\}$. Randomly assign the codewords ${{\mathbf{w}}_{\mathbf{m}}}^n\left( {{s_m}} \right)$ to one of ${2^{nR_m}}$ bins using a uniform distribution. Let $ B_m(t_m) $ denote the set of codeword indices $s_m$ assigned to bin index $t_m$.

\emph{Encoding:} Given a source sequence $X_m^{n}$, the encoder looks for a codeword $W_m^n(s_m)$ such that $\left( {{X_m^n},W_m^n\left( {{s_m}} \right)} \right) \in A_\epsilon^{*\left( n \right)}$. The encoder sends the index of the bin $t_m$ in which $s_m$ belongs.

\emph{Decoding:} The decoder looks for a pair $(W_1^n(s_1),W_2^n(s_2),W_3^n(s_3))$ such that $s_m \in B_m(t_m) $ and $(W_1^n(s_1),W_2^n(s_2), W_3^n(s_3), Y^n) \in A_\epsilon^{*\left( n \right)}$. If the decoder finds a unique triple $(s_1,s_2,s_3)$, he then calculates $\hat{T}^{n}$, where $\hat{T}_i = g(W_{1,i},W_{2,i},W_{3,i},Y_i)$.

\emph{Analysis of the probability of error:}

1. The encoders cannot find the codewords $W_m^n(s_m)$  such that $(X_m^n,W_m^n(s_m)) \in A_\epsilon^{*\left( n \right)}$. The probability of this event is small if
\begin{equation}
\begin{aligned}
  R_m^{'} > I\left( {{X_m},{W_m}} \right) \hfill \\
\end{aligned}
\label{error1}
\end{equation}
2. The pair of sequences  $(X_1^n,W_1^n(s_1)) \in A_\epsilon^{*\left( n \right)}$, $(X_2^n,W_2^n(s_2)) \in A_\epsilon^{*\left( n \right)}$ and $(X_3^n,W_3^n(s_3)) \in A_\epsilon^{*\left( n \right)}$ but the codewords $\{W_1^n(s_1), W_2^n(s_2), W_3^n(s_3)\}$ are not jointly typical with the side information sequences $Y^n$, i.e., $(W_1^n(s_1), W_2^n(s_2), W_3^n(s_3),Y^n) \notin  A_\epsilon^{*\left( n \right)}$. We have assume that 
\begin{equation}
\begin{aligned}
&p\left( {{w_1},{w_2},{w_3},{x_1},{x_2},{x_3},y} \right) \\&= p\left( {{w_1}|{x_1}} \right)p\left( {{w_2}|{x_2}} \right)p\left( {{w_3}|{x_3}} \right)p\left( {{x_1},{x_2},{x_3},y} \right).
\end{aligned}
\label{error2}
\end{equation}
Hence, by the Markov lemma, the probability of this event goes to zero if $n$ is large enough.

3. There exists another $s^{'}$ with the same bin index that is jointly typical with the side information sequences.
The correct codeword indices are denoted by $s_1$ , $s_2$ and $s_3$. We first consider the situation where the codeword index $s_1$ is in error. The probability that a randomly chosen $W_1^n(s^{'}_1)$ is jointly typical with $(W_2^n(s_2),W_3^n(s_3), Y^n)$ can be bounded as
\begin{equation}
\begin{small}
\begin{aligned}
&\Pr \left\{ {\left( {W_1^n\left( {s_1^{'}} \right),W_2^n\left( {{s_2}} \right), W_3^n\left( {{s_3}} \right),{Y^n}} \right) \in A_\epsilon^{*\left( n \right)}} \right\}\\& \leqslant {2^{ - n\left( I{\left( {{W_1};{W_2},{W_3},Y} \right) - 3\epsilon} \right)}}.
\end{aligned}
\label{error3}
\end{small}
\end{equation}
The probability of this error event is bounded by the number of codewords in the bin  $t_1$ times the probability of joint typicality
\begin{equation}
\begin{small}
\begin{aligned}
  &\Pr \left\{ \exists s_1^{'} \in {B_1}\left( {{t_1}} \right),s_1^{'} \ne {s_1}:\right.\\
  &\left.\left( {W_1^n\left( {s_1^{'}} \right),W_2^n\left( {{s_2}} \right),W_3^n\left( {{s_3}} \right),{Y^n}} \right) \in A_\epsilon^{*\left( n \right)} \right\} \hfill \\&
   \leqslant \sum\limits_{s_1^{'} \ne {s_1},\atop s_1^{'} \in {B_1}\left( {{t_1}} \right)} {\Pr \left\{ {\left( {W_1^n\left( {s_1^{'}} \right),W_2^n\left( {{s_2}} \right),W_3^n\left( {{s_3}} \right),{Y^n}} \right) \in A_\epsilon^{*\left( n \right)}} \right\}}  \hfill \\
   &\leqslant {2^{n\left( {{R_1} - R_1^{'}} \right)}}{2^{ - n\left( {I\left( {{W_1};{W_2},{W_3},Y} \right) - 3\epsilon} \right)}}.
\end{aligned}
\label{error4}
\end{small}
\end{equation}
Similarly, the probability that the codeword index $s_2$ or $s_3$ is in error can be bounded by
\begin{equation}
\begin{aligned}
  &\Pr \left\{ \exists s_2^{'} \in {B_2}\left( {{t_2}} \right),s_2^{'} \ne {s_2}:\right.\\
  &\left.\left( {W_1^n\left( {s_1} \right),W_2^n\left( {{s_2^{'}}} \right),W_3^n\left( {{s_3}} \right),{Y^n}} \right) \in A_\epsilon^{*\left( n \right)} \right\} \hfill \\&
\leqslant {2^{n\left( {{R_2} - R_2^{'}} \right)}}{2^{ - n\left( {I\left( {{W_2};{W_1},{W_3},Y} \right) - 3\epsilon} \right)}},\\
  &\Pr \left\{ \exists s_3^{'} \in {B_3}\left( {{t_3}} \right),s_3^{'} \ne {s_3}:\right.\\
  &\left.\left( {W_1^n\left( {s_1} \right),W_2^n\left( {{s_2}} \right),W_3^n\left( {{s_3^{'}}} \right),{Y^n}} \right) \in A_\epsilon^{*\left( n \right)} \right\} \hfill \\&
\leqslant {2^{n\left( {{R_3} - R_3^{'}} \right)}}{2^{ - n\left( {I\left( {{W_2};{W_1},{W_3},Y} \right) - 3\epsilon} \right)}}.
\end{aligned}
\label{error4_1}
\end{equation}

We then consider the case that two of the three codeword indices are in error. The probability that the randomly chosen $W_1^n(s^{'}_1)$ and $W_2^n(s^{'}_2)$ are jointly typical with $(W_3^n(s_3), Y^n)$ can be bounded as 
\begin{equation}
\begin{aligned}
  &\Pr \left\{ {\left( {W_1^n\left( {s_1^{'}} \right),W_2^n\left( {s_2^{'}} \right),W_2^n\left( {{s_3}} \right),Y^n} \right) \in A_\epsilon^{*\left( n \right)}} \right\} \hfill \\&
   = \sum\limits_{\left( {{W_1},{W_2},{W_3},Y} \right) \in A_\epsilon^{*\left( n \right)}} {p\left( {w_1^n} \right)p\left( {w_2^n} \right)p\left( {w_3^n,y^n} \right)}  \hfill \\&
   \leqslant  {2^{ - n\left( {I\left( {{W_1};{W_2},{W_3},Y} \right) + I\left( {{W_2};{W_1},{W_3},Y} \right) - I\left( {{W_1};{W_2}|{W_3},Y} \right) - 4\epsilon} \right)}} \hfill.
\end{aligned}
\label{error5}
\end{equation}
Hence, the error probability can be bounded as
\begin{equation}
\begin{aligned}
  &\Pr \left\{ \exists s_1^{'} \in {B_1}\left( {{t_1}} \right),s_1^{'} \ne {s_1},\exists s_2^{'} \in {B_2}\left( {{t_2}} \right),s_2^{'} \ne {s_2}:\right.\\
  &\left. \left( {W_1^n\left( {s_1^{'}} \right),W_2^n\left( {s_2^{'}} \right),W_2^n\left( {{s_3}} \right),y^n} \right) \in A_\epsilon^{*\left( n \right)} \right\} \hfill \\
  &\leqslant  {2^{n( {{R_1} - R_1^{'} + {R_2} - R_2^{'}} )}}\\
   &\times 2^{ - n\left( {I\left( {{W_1};{W_2},{W_3},Y} \right) + I\left( {{W_2};{W_1},{W_3},Y} \right)  - I\left( {{W_1};{W_2}|{W_3},Y} \right) - 4\epsilon} \right)} \hfill .
\end{aligned}
\label{error6}
\end{equation}
Similarly, we can obtain the probability that the codeword indices $(s_1,s_3)$ or $(s_2,s_3)$  are in error, which we omit here.

For the case where all the codeword indices $s_1$, $s_2$ and $s_3$ are in error. The probability that the randomly chosen $W_1^n(s^{'}_1)$, $W_2^n(s^{'}_2)$ and $W_3^n(s^{'}_3)$  are jointly typical with $Y^n$ can be bounded as
\begin{equation}
\begin{small}
\begin{aligned}
  &\Pr \left\{ {\left( {W_1^n\left( {s_1^{'}} \right),W_2^n\left( {s_2^{'}} \right),W_2^n\left( {s_3^{'}} \right),Y^n} \right) \in A_\epsilon^{*\left( n \right)}} \right\} \hfill \\&
   = \sum\limits_{\left( {{W_1},{W_2},{W_3},Y} \right) \in A_\epsilon^{*\left( n \right)}} {p\left( {w_1^n} \right)p\left( {w_2^n} \right)p\left( {w_3^n} \right)p\left( {Y^n} \right)}  \hfill \\&
   \leqslant  2^{ - n\left( I\left( {{W_1};{W_2},{W_3},Y} \right) + I\left( {{W_2};{W_1},{W_3},Y} \right)\right)}\\
   &\times 2^{ -n\left( I\left( {{W_3};{W_1},{W_2},Y} \right) - I\left( {{W_1};{W_2}|{W_3},Y} \right) - I\left( {{W_1},{W_2};{W_3}|Y} \right) - 5\epsilon \right)} \hfill.
\end{aligned}
\end{small}
\label{error7}
\end{equation}

The probability of the above error events goes to 0 when
\begin{equation*}
\begin{footnotesize}
\begin{aligned}
  R_1^{'} - {R_1} &\leqslant I\left( {{W_1};{W_2},{W_3},Y} \right) \hfill \\ &...
\end{aligned}
\label{error8}
\end{footnotesize}
\end{equation*}
\begin{equation}
\begin{footnotesize}
\begin{aligned}
  R_1^{'} - {R_1} + R_2^{'} - {R_2} + R_3^{'} - {R_3} &\leqslant I\left( {{W_1};{W_2},{W_3},Y} \right) \\&+ I\left( {{W_2};{W_1},{W_3},Y} \right)\\& + I\left( {{W_3};{W_1},{W_2},Y} \right) \\&- I\left( {{W_1};{W_2}|{W_3},Y} \right)\\& - I\left( {{W_1},{W_2};{W_3}|Y} \right).
\end{aligned}
\label{error8}
\end{footnotesize}
\end{equation}

Therefore, \eqref{achieve region11} can be obtained by combining (\ref{error1}) and (\ref{error8}).

If $(s_1, s_2, s_3)$ are correctly decoded, we have $(X_1^n, X_2^n, X_3^n, W_1^n(s_1), W_2^n(s_2), W_3^n(s_3), Y^n) \in A_\epsilon^{*\left( n \right)}$. Therefore, the empirical joint distribution is close to the distribution $p\left( {{w_1}|{x_1}} \right)p\left( {{w_2}|{x_2}} \right)p\left( {{w_3}|{x_3}} \right)p\left( {{x_1},{x_2},{x_3},y} \right)$ that achieves distortion $D$.

\section*{Appendix B}
Here we provide the rigorous proof of Theorem 3.
Define a series of encoders $f_m: \mathcal{X}_m^n \rightarrow\left\{1, \ldots, 2^{n R_m}\right\}$ and decoders $g: \prod_{m \in \mathcal{M}}\left\{1, \ldots, 2^{n R_m}\right\} \times \mathcal{Y}^n \rightarrow \hat{\mathcal{T}}^n$ that achieving the given distortion $D$. We can derive the following inequalities
\begin{subequations} \label{converse1}
\begin{small}
\begin{align}
&nR_1 \nonumber\\&\geq H(f_1(X_1^n))\nonumber\\
&\geq  H(f_1(X_1^n)|f_2(X_2^n),f_3(X_3^n), Y^n) \label{converse1a}\\
&\geq  H(f_1(X_1^n)|f_2(X_2^n),f_3(X_3^n), Y^n) \nonumber\\ 
&- H(f_1(X_1^n)|f_2(X_2^n),f_3(X_3^n), Y^n, X_1^n, X_2^n, X_3^n)\nonumber\\
&=  I(X_1^n, X_2^n,X_3^n; f_1(X_1^n)|f_2(X_2^n), f_3(X_3^n),Y^n) \label{converse1b}\\
&= \sum\limits_{i=1}^n I(X_{1,i}, X_{2,i}, X_{3,i}; f_1(X_1^n) | f_2(X_2^n),f_3(X_3^n), \nonumber\\&Y^n, X_{1,1}^{i-1}, X_{2,1}^{i-1}, X_{3,1}^{i-1}) \label{converse1c}\\
&=\sum\limits_{i=1}^n H(X_{1,i}, X_{2,i} | f_2(X_2^n), Y^n, X_{1,1}^{i-1}, X_{2,1}^{i-1}) \nonumber\\
&- H(X_{1,i}, X_{2,i} | f_1(X_1^n) , f_2(X_2^n), Y^n, X_{1,1}^{i-1}, X_{2,1}^{i-1} ) \label{converse1d}\\
&= \sum\limits_{i=1}^n H(X_{1,i}, X_{2,i} | W_{2,i}, Y_i)- H(X_{1,i}, X_{2,i} | W_{1,i}, W_{2,i}, Y_i) \label{converse1e}\\
&= \sum\limits_{i=1}^n I(X_{1,i}, X_{2,i}; W_{1,i} | W_{2,i}, Y_i)\nonumber,
\end{align}
\end{small}
\end{subequations}
where \eqref{converse1a} follows from the fact that conditioning reduces entropy, \eqref{converse1b} and \eqref{converse1d}  are obtained by the definition of conditional mutual information and \eqref{converse1c} is the chain rule of mutual information. In \eqref{converse1e}, we let ${W_{1,i}} = \left( {{f_1(X_1^n)},X_{1,1}^{i - 1},X_{2,1}^{i - 1},Y_1^{i - 1},}{Y_{i + 1}^n} \right)$, and ${W_{2,i}} = \left( {{f_2(X_2^n)},X_{1,1}^{i - 1},X_{2,1}^{i - 1},Y_1^{i - 1},}{Y_{i + 1}^n} \right)$. 
Similarly we have
\begin{equation} 
\begin{aligned}
nR_2\geq \sum\limits_{i=1}^n I(X_{1,i}, X_{2,i},X_{3,i}; W_{2,i} | W_{1,i},W_{3,i}, Y_i),\\
nR_3\geq \sum\limits_{i=1}^n I(X_{1,i}, X_{2,i},X_{3,i}; W_{3,i} | W_{2,i},W_{3,i}, Y_i).
\end{aligned}
\label{converse1}
\end{equation}

The rigorous proof of the sum rate part will be provided in a longer version.

\clearpage
\bibliographystyle{IEEEtran}
\bibliography{ref}

\end{document}